\documentclass[submission, LectureNotes]{SciPost}
\usepackage{lineno}
\usepackage{graphicx}
\usepackage{hyperref}
\usepackage{color}
\usepackage{pythonhighlight}
\newcommand{\br}{{\bf r}}
\newcommand{\R}{{\bf R}}
\newcommand\makemargintip[1]{\textcolor{scipostdeepblue}{\fbox{\color{black}#1}}%
  \marginpar{\includegraphics[scale=0.3]{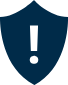}}}

\begin{document}

\begin{center}{\Large \textbf{
A Practical Introduction to Density Functional Theory
}}\end{center}

\begin{center}
L. Rademaker\textsuperscript{1*}
\end{center}

\begin{center}
{\bf 1} Department of Theoretical Physics, University of Geneva, 1211 Geneva, Switzerland
\\
* louk.rademaker@gmail.com
\end{center}

\begin{center}
\today
\end{center}


\section*{Abstract}
{\bf
These lecture notes contain a brief practical introduction to doing density functional theory calculations for crystals using the open source Quantum Espresso software. The level is aimed at graduate students who are studying condensed matter or solid state physics, either theoretical or experimental.
}

\vspace{10pt}
\noindent\rule{\textwidth}{1pt}
\tableofcontents\thispagestyle{fancy}
\noindent\rule{\textwidth}{1pt}
\vspace{10pt}

\section*{Preface}

These notes were developed for a short 4-hour course on density functional theory (DFT) given in March 2020 at the University of Geneva. The lectures were the first instalment of the 'ToolBoX' series of lectures on different modern tools used in condensed matter physics. We require basic knowledge of solid state physics and quantum mechanics - most importantly band theory. Other than that, the idea is to be very {\em practical}. Making your hands dirty with an actual calculation is, after all, the most reliable method of learning something.

Below, we will first give a speeding review of what DFT is in Sec.~\ref{sec:intro}. How to prepare a DFT calculation is discussed in Sec.~\ref{Sec:Guide}. The remaining sections contain three hands-on examples of real-world materials whose properties can teach us something about how to do DFT: silicon, graphene and WSe$_2$. You are supposed to work out these examples yourself, but lazy students can download the set of input files and pseudopotentials here: \url{http://loukrademaker.nl/dft_files/}.This course is nowhere intended to make you a world-expert; rather, it gives you a good head start and in Sec.~\ref{Sec:Learn} we mention some further places you can learn about DFT.

\section{Introduction: What is Density Functional Theory?}
\label{sec:intro}

Any material on earth, whether in crystals, amorphous solids, molecules or yourself, consists of nothing else than a bunch of atoms, ions and electrons bound together by electric forces. All these possible forms of matter can be explained by virtue of one simple equation: the many-particle Schr\"{o}dinger equation,
\begin{equation}
	i \hbar \frac{\partial}{\partial t} \Phi( \br ; t) = 
	\left( - \sum_{i}^N \frac{\hbar^2}{2m_i} \frac{\partial^2}{\partial \br_i^2}
	+ \sum_{i<j}^N \frac{e^2 Z_i Z_j }{ | \br_i - \br_j| } \right)  \Phi( \br ; t).
	\label{Eq:ManyPSchrEe}
\end{equation}
Here $\Phi(\br ; t)$ is the many-body wavefunction for $N$ particles, where each particle has its own mass $m_i$, charge $Z_i$ and position $\br_i$. The only interaction is the Coulomb interaction $e^2/r$. 

Despite its apparent simplicity, Eq.~\eqref{Eq:ManyPSchrEe} is notoriously difficult to solve. This is where density functional theory (DFT) comes in. Using a set of reasonable physical approximations we can simplify the many-particle Schr\"{o}dinger equation to something that we can actually solve numerically. 

\subsection{Born-Oppenheimer approximation}
The first approximation arises from the physical problem we want to study: the ground state of a collection of interacting ions and electrons. Because even the lightest ion is more than a thousand times heavier than an electron, we will forget about the dynamics of the ions all-together. This is known as the {\em Born-Oppenheimer approximation}. 
We then write the time-independent Schr\"{o}dinger equation for a collection of $N$ electrons {\em subject to the electric potential created by the fixed ions},
\begin{equation}
	\left( \sum_{i}^N \left( - \frac{\hbar^2}{2m} \frac{\partial^2}{\partial \br_i^2}
	+ V(\br_i) \right)
	 + \sum_{i<j}^N \frac{e^2}{|\br_i - \br_j|} \right) \Psi(\br) = E_0  \Psi( \br )
	\label{Eq:BornOppenheimer}
\end{equation}
where $\br_i$ are the positions of the electrons. The potential $V(\br_i)$ is created by the charged ions,
\begin{equation}
	V(\br_i) = - \sum_j \frac{e^2 Z_j}{| \br_i - \R_j|}
\end{equation}
where $\R$ is the (static) positions of the ions and $Z_j$ their charge. 
Note that the above Hamiltonian -- the left hand side of Eq.~\eqref{Eq:BornOppenheimer} -- contains three terms: the kinetic energy ($T$), the potential energy ($V$) and the interaction energy ($U$). 

The electronic density is obtained by integrating out all electron degrees of freedom except one,
\begin{equation}
	n(\br) = \int d^3 \br_2 \cdots d^3 \br_N
		\left| \Psi (\br_1 \cdots \br_n) \right|^2.
\end{equation}
The total potential energy $V$ is just given by the integral over the potential $V(\br)$ times the density,
\begin{equation}
	V = \int d^3 \br V(\br) n(\br).
\end{equation}

\subsection{Hohenberg-Kohn theory}

Assume we found a solution of Eq.~\eqref{Eq:BornOppenheimer}, with ground state energy $E_0$ and a certain electronic density $n(\br)$. The strength of the Coulomb interaction and the mass of an electron are constants of nature, so the only input that can possibly influence the electronic density $n(\br)$ and the energy $E_0$ of our ground state is our choice of potential $V(\br)$. In other words, the ground state energy is a {\em functional} of the input potential,
\begin{equation}
	E_0 [V(\br)] = \mathcal{F}_E[ V(\br) ]
	\label{Eq:E0}
\end{equation}
A functional is nothing else than a function whose input is another function; in this case the functional $\mathcal{F}$ takes as input the electric potential generated by the ions and outputs the ground state energy based on Eq.~\eqref{Eq:BornOppenheimer}. 

At first this results seems counterintuitive. After all, the ground state energy clearly contains the kinetic energy $T$, the interaction energy $U$ and the potential energy. Only the latter term {\em explicitly} depends on the potential. We can thus write the ground state energy in terms of a separate functional for the kinetic and interaction energy, and the potential energy
\begin{equation}
	E  [n(\br)]= \mathcal{F}_E' [ V(\br) ] + \int d^3\br V(\br)  n(\br)
	\label{Eq:E02}
\end{equation}

Hohenberg and Kohn\cite{Kohn:1964ut} came to the elegant insight that the {\em potential $V(\br)$ and electronic density $n(\br)$ are conjugate variables}. Other conjugate variables you may know are for example pressure and volume in thermodynamics or momentum and position in classical physics. The fact that the potential $V(\br)$ and the density $n(\br)$ are conjugate means you can equally well describe any solution of Eq.~\eqref{Eq:BornOppenheimer} using the potential {\em or} the density. 

Formally known as a Legendre transform (in the same way you go from the Hamiltonian to the Lagrangian formulation of classical mechanics), we can change the functional of Eq.~\ref{Eq:E02} to depend on the density $n(\br)$ rather than the potential $V(\br)$. This is the {\em Hohenberg-Kohn theorem}: there exists a {\em universal} functional of electronic density, $\mathcal{F}[n(\br)]$, such that for the correct density $n(\br)$ it provides the ground state energy of Eq.~\eqref{Eq:BornOppenheimer},
\begin{equation}
	E [n(\br)] = \mathcal{F}[n(\br)] +  \int d^3\br V(\br)  n(\br).
\end{equation}
Knowing this functional, for any given potential $V(\br)$ we minimize the right hand side by checking all possible electronic density distributions.

There are only two minor problems. We don't know what this functional looks like. And even if we did, we don't know how to find the right electronic density.

\subsection{Approximating the functional}
\label{Sec:ApproxFunc}

The unknown functional $\mathcal{F}[n(\br)]$ should describe the kinetic and interaction energy of a system described by Eq.~\ref{Eq:BornOppenheimer}. Even though we cannot find its exact shape, we can look at its shape in some limiting cases that we can solve.

We know that a free homogeneous electron gas with density $n$ has a ground state energy of
\begin{equation}
	E_0 = \frac{3 \hbar^2 \left( 3 \pi^2 \right)^{2/3}}{10 m} n_0^{5/3}.
	\label{Eq:FHEG}
\end{equation}
For a slowly varying electronic density, we can approximate the kinetic energy contribution to the full functional $\mathcal{F}[n(\br)]$ as the energy of Eq.~\eqref{Eq:FHEG} evaluated at each point separately,
\begin{equation}
	\mathcal{T}_0 [n(\br)] =
	 \frac{3 \hbar^2 \left( 3 \pi^2 \right)^{2/3}}{10 m} \int  d^3\br ( n(\br) )^{5/3}.
	 \label{Eq:T0}
\end{equation}

Furthermore, we know from perturbation theory that the lowest order energy contribution from Coulomb interactions is given by the Hartree term,
\begin{equation}
	\mathcal{U}_{\mathrm{H}} [n(\br)] = 
		\frac{e^2}{2} \int d^3\br d^3\br' \, \frac{n(\br) n(\br')}{|\br - \br'|}.
\end{equation}

It is natural to write out the full functional as containing the homogeneous electron gas term and the Hartree term. The remaining terms, though still unknown, should be small. This unknown part is conventionally called the {\em exchange-correlation potential} $E_{xc}[n(\br)]$. The full Hohenberg-Kohn functional, including the potential energy, is thus
\begin{equation}
	\mathcal{E}_{\mathrm{HK}}[n(\br)] = 
		\mathcal{T}_0 [n(\br)]
		 + \int d^3 \br V(r) n(\br)
		 + \mathcal{U}_{\mathrm{H}} [n(\br)]
		 + E_{xc} [ n(\br)].
	\label{Eq:HKFunctional}
\end{equation}
We will later discuss some general choices of exchange-correlation functionals in Sec.~\ref{Sec:Functionals}.

\subsection{Kohn-Sham equation}

We replaced an intractable problem (solving Eq.~\eqref{Eq:BornOppenheimer}) with the task of minimizing an unknown functional $\mathcal{F}[n(\br)]$ over infinitely many possible electronic densities $n(\br)$. In the previous section we already gave some first suggestions for the functional. But once we found it, how to find the right electronic density $n(\br)$?

Because the correct density minimizes the functional, we can find the functional by setting it's derivative to zero,
\begin{equation}
	\frac{\delta \mathcal{F}[n(\br)] }{\delta n(\br)} = 0.
\end{equation}
Using the functional Eq.~\eqref{Eq:HKFunctional}, we write out\footnote{There is a small subtlety included in this equation: the kinetic component of the functional $\mathcal{T} [n(\br)]$ should not be the one obtained for the free homogeneous noninteracting electron gas of Eq.~\eqref{Eq:T0}, but the one for a noninteracting gas subject to the Kohn-Sham potential.}
\begin{equation}
	\frac{\delta \mathcal{T}[n(\br)] }{\delta n(\br)}
	+ V(\br)
	+ \int \frac{n(\br')}{|\br-\br'|}d^3 \br'
	+ \frac{\delta E_{xc} [n(\br)] }{\delta n(\br)}
	=0.
\end{equation}
The idea of Kohn and Sham\cite{Kohn:1965js} was to treat this as if it is a {\em single-particle problem}. The first term represents the kinetic energy, and the remaining terms form the {\em Kohn-Sham potential}
\begin{equation}
	V_{\mathrm{KS}}(\br) = V(\br)
	+ \int \frac{n(\br')}{|\br-\br'|}d^3 \br'
	+ \frac{\delta E_{xc} [n(\br)] }{\delta n(\br)}.
	\label{Eq:KSpotential}
\end{equation}
The {\em Kohn-Sham equation} is the single-particle Schr\"{o}dinger equation with the potential given by Eq.~\eqref{Eq:KSpotential},
\begin{equation}
	\left( - \frac{\hbar^2}{2m} \frac{\partial^2}{\partial \br^2} + V_{\mathrm{KS}}(\br) \right) \psi_i(\br) = \epsilon_i \psi_i (\br).
	\label{Eq:KSeq}
\end{equation}
We solve these equations numerically, which is tractable because it's just a linear differential equation. The electronic density is obtained by occupying the $N$ solutions  $\psi_i(\br)$ with the lowest energy, \begin{equation}
	n(\br) = \sum_{i=1}^N |\psi_i (\br) |^2.
\end{equation}
Now the electronic density obtained this way can be used to calculate a new Kohn-Sham potential following Eq.~\eqref{Eq:KSpotential}. We continue this iterative procedure until we reach convergence.

A final comment is in order: in the above derivation we completely ignored the spin of electrons. Of course, real electrons have spin so that you need that degree of freedom as well. This does not change anything fundamental about how to use the Kohn-Sham equation.

\subsection{Limitations}

We have reached the end-goal: using the Born-Oppenheimer  approximation, with an appropriate choice of functional, we use the Kohn-Sham equations to find the ground state energy and electronic density of a system of interacting electrons and ions. This combination of approximations and techniques is called {\em density functional theory} (DFT).

Despite is sometimes shaky assumptions, DFT turned out to be a resounding success. A large majority of crystalline materials, many molecules and molecular structures have been explained using DFT. Walter Kohn -- the man who was involved in both the Hohenberg-Kohn theory and the Kohn-Sham equations -- received the Nobel Prize for DFT in 1998. Mainly because of this success, I assume, you want to learn DFT in this ToolBoX.

However, let me briefly shed some clouds over DFT's success. 

\begin{itemize}
	\item A major limitation of DFT is that it is impossible to tell you the size of the errors that exist due to the assumptions. If you get a self-consistent solution, that is nice, but only a comparison with the experimental system will tell you whether it is a good solution.
	\item The electronic properties of many materials can be described using band theory, meaning for every quasimomentum ${\bf k}$ we have a set of energies $\epsilon_n ({\bf k})$. The solutions of the Kohn-Sham equation Eq.~\eqref{Eq:KSeq} are commonly interpreted as these electronic bands. However, it is important to bear in mind that in principle there is no connection to the {\em actual} electronic energy levels in a material. It just turns out that, in many materials, the Kohn-Sham energies happens to be a good approximation.
	\item As a corollary to the previous limitation: when using DFT for an insulator or semiconductor, you can also compute the Kohn-Sham energy gap between the highest occupied and lowest unoccupied state. This is {\em not} the actual gap of the semiconductor or insulator. In practice, it turns out that DFT typically underestimates the real gap.
	\item Note that even if we had the exact functional, solving the corresponding Kohn-Sham equations would not give you the exact solution for the ground state energy.
	\item Because the Kohn-Sham equations describe non-interacting electrons, many materials with strong correlations cannot be described using DFT. In general, this is true for materials with partially filled $d$ or $f$-orbitals; or materials with localized electrons. In particular, applying DFT to the class of high-temperature superconductors such as cuprates, pnictides, and heavy fermions is relatively unsuccessful. 
	\item Materials with ground state degeneracy -- for example in the case of spontaneous symmetry breaking -- are known to be difficult to compute using DFT.
	\item There is some subtlety involved in choosing the right functional. Commonly, functionals become widely accepted because of a good overlap with experiments.
\end{itemize}

A good book to learn more about the theoretical side of DFT is Ref.~\cite{giuliani2005quantum}.

\section{A Practical Guide}
\label{Sec:Guide}
DFT as introduced in the last section is a numerical technique -- there are no analytical expansions or solutions. And like many modern numerical techniques, it is better to use an existing developed code than to write your own. There are numerous DFT software packages, both open source and payed. In this section we will outline the most important choices we need to make in order to start running some code.

\subsection{Wavefunction basis sets}

The core of any code consists of computing the solutions to the Kohn-Sham equation. Because this is a linear differential equation, we need to choose a {\em basis} over which we can expand the Kohn-Sham equation. In this way, we transformed the continuum differential equation into a matrix equation, for which there are many known techniques for diagonalizing it.

The two most popular choices of basis are {\em plane waves} (PW) and {\em Gaussian type orbitals} (GTO). If you are interested in the structure of molecules, the most logical basis set is GTO where you take a certain set of polynomials multiplied by a Gaussian envelope, to make sure the electronic density remains close to the ions.

For crystals, on the other hand, the most logical basis set is plane waves $\psi({\bf k}) = e^{i {\bf k} \cdot {\bf r}}$ in a box with periodic boundary conditions. Because this course is aimed at condensed matter physicists, we will use a plane-wave code.

A full list of existing DFT codes, with their preferred basis set, is maintained on Wikipedia\cite{wiki:DFTlist}.

\subsection{Which functional?}
\label{Sec:Functionals}

In Sec.~\ref{Sec:ApproxFunc} we introduced some basic ideas regarding the precise shape of the functional. We separated the kinetic energy of a noninteracting gas and the Hartree-Fock interaction energy from the exchange-correlation functional. Here we will discuss in more detail some possible exchange-correlation functionals that have been proposed, without any attempt at being encyclopedic. 

It is known -- see for example the textbook \cite{Mahan:2000wd}, chapter 5 -- that the energy of the homogeneous electron gas can be expanded in powers of the Fermi momentum $k_F \sim n^{1/3}$. Kohn and Sham\cite{Kohn:1965js} suggested to use these analytical results for the exchange-correlation functional, which is now known as the {\em local density approximation} (LDA).

A natural next step is to have a functional that not only depends on the density $n(\br)$ but also on its derivative $\nabla n(\br)$. Such functionals are known as generalized gradient approximations (GGA).\cite{PhysRevB.46.6671} A popular version of a GGA functional is the {\em Perdew–Burke-Ernzerhof functional} (PBE) \cite{PhysRevLett.77.3865}  - its publication is cited more than 100000 times! In this work we will use the PBE functional, since it reproduces experimental band-structures relatively accurate.

\subsection{Pseudopotentials}

Are we ready to start computing? Well, not yet. If we are interested in doing DFT for a crystal, we would give our code the size of the unit cell, the type of atoms and their positions. For example, silicon has an {\em fcc} crystal structure with two atoms per unit cell. Silicon itself is element number 14, which means there are 28 electrons per unit cell. However, both physically and numerically it is nonsensical to include all 28 electrons in our calculation.

Physically speaking, the electronic configuration of silicon is Ne 3s2 3p2. The core electrons, given by the electronic configuration of neon, are completely irrelevant in the physics of binding a silicon crystal. Numerically, the fact that a bare atomic core has a diverging potential $1/r$ creates a lot problems. Both these problems can be solved by putting a {\em pseudopotential} at the position of the silicon atoms. A pseudopotential is a smeared-out potential that includes the charge of the physically irrelevant core electrons, that is therefore more numerically stable than the diverging $1/r$ potential of a bare atomic core. Using a pseudopotential for silicon, for example, means you now do DFT with only the four outermost electrons, known as the {\em valence} electrons. 

Like with the choice of functional, there are many different ways to compute a pseudopotential. In fact, most pseudopotentials are tailored to work with certain functionals, so in these notes we will use pseudopotentials that work well with the PBE functional.

\subsection{Quantum ESPRESSO}

Now we are ready to select a DFT implementation. In these notes we will use the open-source plane-wave DFT code {\sc Quantum ESPRESSO} (\url{http://www.quantum-espresso.org/}). Its development was started in Trieste, Italy, but has by now many contributors from all around the world. If you ever use {\sc Quantum ESPRESSO} scientifically, make sure you explicitly acknowledge the code and cite their original journal publications\cite{QE-2009,QE-2017}.

Our first task is to {\em install} {\sc Quantum ESPRESSO} on your own computer. The simulations we will do in these lecture notes are light enough that they can be done on a standard laptop. 

The full source package of {\sc Quantum ESPRESSO} can be downloaded from
\begin{center}
\url{https://www.quantum-espresso.org/download}
\end{center}
If you are comfortable doing so, you can install {\sc Quantum ESPRESSO} using the source package. Otherwise, on the above webpage you will also find links for stable binaries for typical platforms such as Windows. For Mac OS X I recommend using MacPorts (\url{https://www.macports.org/}), which has a port called \verb+quantum-espresso+. It automatically takes care of dependencies, such as OpenMPI and Fortran libraries.

\subsection{Materials Cloud}
We will use the pseudopotential libraries collected by {\em Materials Cloud},\cite{Lejaeghere:2016jx,Prandini:2018wx} an online tool and repository for doing DFT calculations developed by the EPFL and the ETH. On their website, they have collected various pseudopotentials, benchmarked them, and selected for each element the best choice of pseudopotential.

\begin{figure}[t]
\begin{center}
	\includegraphics[width=0.8\textwidth]{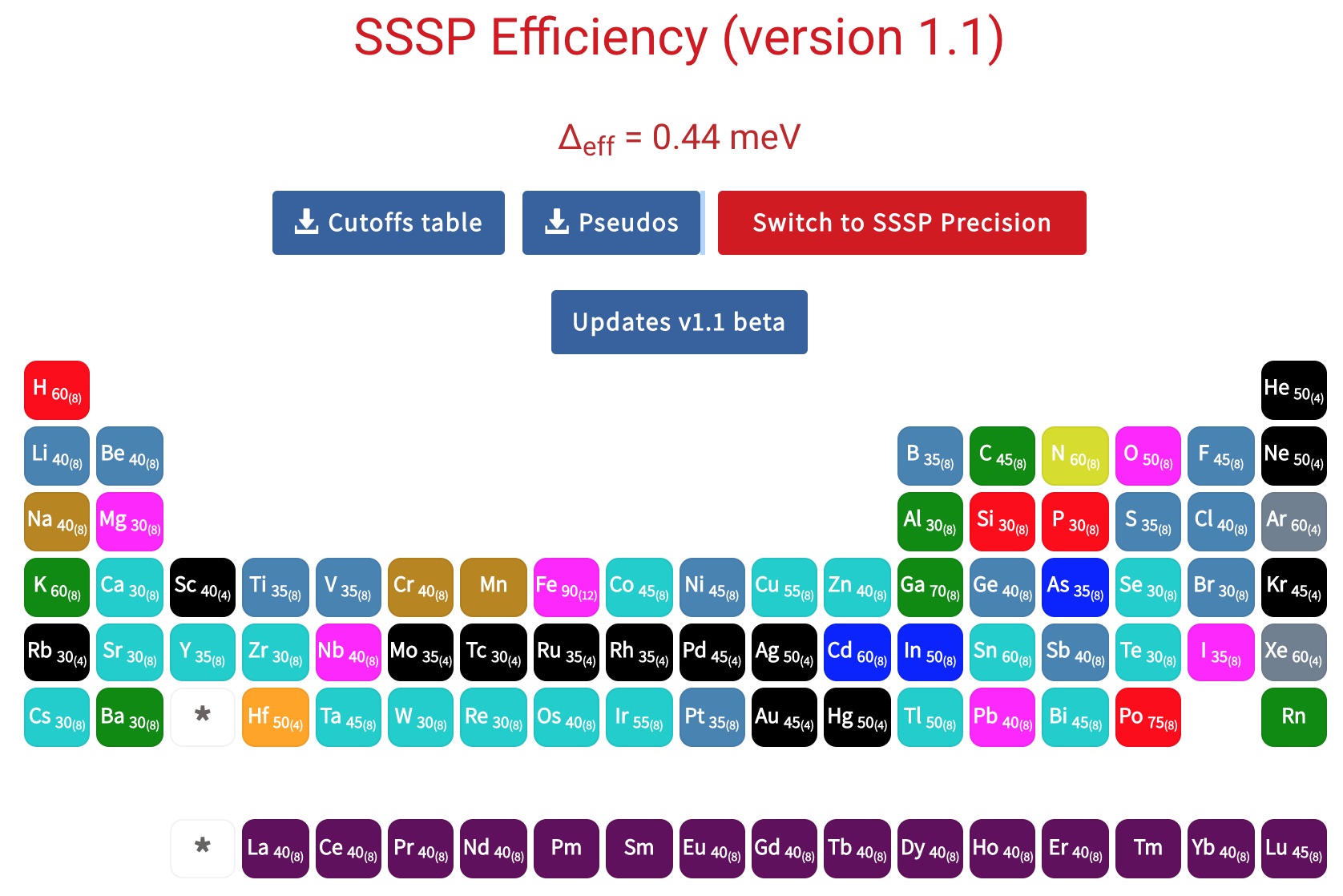}
\end{center}
	\caption{The {\em Standard solid-state pseudopotentials library} from Materials Cloud\cite{Lejaeghere:2016jx,Prandini:2018wx} contains for all elements a choice of pseudopotential, which can be downloaded on \url{https://www.materialscloud.org/}.}
	\label{Fig:SSSP}
\end{figure}

We will download their collection of pseudopotentials, which are all computed for the PBE functional.
\begin{enumerate}
\item Go to the website \url{https://www.materialscloud.org/}, and navigate to the {\em Discover} page using the top menu.
\item Click on {\em Standard solid-state pseudopotentials (SSSP)}.
\item You will see a periodic table (see Fig.~\ref{Fig:SSSP}). By clicking on the button that reads {\em Pseudo}, you will download a \verb+tar.gz+ file containing all the pseudopotentials.
\item Once you downloaded the library, look into the folder. You will see files with names like \verb+Si.pbe-n-rrkjus_psl.1.0.0.UPF+ and \verb+C.pbe-n-kjpaw_psl.1.0.0.UPF+. These are the pseudopotentials for silicon and carbon, respectively. Directly after that, you can see that these pseudopotentials are computed for the PBE functional.
\end{enumerate}

\section{Example 1: Silicon}
\label{Sec:Silicon}

Silicon is one of the most abundant materials on earth, and one of the most used in modern technology. It is therefore logical that we start with silicon as the first material we are going to study. 

To start off, create a folder where we will do all of our calculations, with three subfolders: \verb+silicon+, \verb+out+ and \verb+pseudo+.
Then move the pseudopotential file for silicon that we downloaded from Materials Cloud into the \verb+pseudo+ directory.

\subsection{Self-consistent field}

\subsubsection{Writing the input file}
{\sc Quantum ESPRESSO} works with input files. These are text-only files that contain all the parameters that you want to give to your code. We will now build together an input file for computing the ground state of a silicon crystal. Note that all possible parameters are summarized in the file \verb+INPUT_PW.txt+ or \verb+INPUT_PW.html+ that came with the source package.

So let's get coding! Open your favorite plain text editor, and create a new file called \verb+silicon.scf.in+. The input file we will create is structured with {\em cards}, of the form  \verb+&NAME ... /+.

The first card named \verb+CONTROL+ describes calculation parameters.
\begin{python}
&CONTROL
  calculation = 'scf'
  prefix = 'silicon'
  outdir = '../out/'
  pseudo_dir = '../pseudo/'
  tprnfor = .true.
  verbosity = 'high'
/
\end{python}
\begin{itemize}
	\item \verb+calculation+ decides what calculation we will do. Choose \verb+'scf'+, which is short for self-consistent field calculation. It is the default option of {\sc Quantum ESPRESSO}. In the scf-mode, the self-consistent loop described in Sec.~\ref{sec:intro} is done to find the ground state energy and the electronic density.
	\item \verb+prefix+ is the name of your set of calculations. This allows you to later use the output of this calculation in further calculations, for example to obtain the band structure.
	\item \verb+outdir+ is the directory where all the detailed output files will be put, in files of the form \verb+prefix.xml+ and \verb+prefix.save+, with the exception of the command line output.
	\item \verb+pseudo_dir+ is the directory where you put the pseudopotentials.
	\item The last two flags are optional. We set \verb+tprnfor+ true, which means the code will calculate and output the forces acting on the ions. This is a useful way to check that your proposed crystal structure is stable. We also set \verb+verbosity+ to high, which means the output file will contain a lot of extra information, such as the calculated symmetry representations of the crystal. I can really advise to run both with high and low verbosity and to compare the output files.
\end{itemize}

The next card describes the system we are studying. What kind of lattice, how many atoms, and what are the cut-offs for our plane wave basis.
\begin{python}
&SYSTEM
  ibrav = 2
  celldm(1) = 10.2
  nat = 2
  ntyp = 1
  occupations = 'fixed'
  ecutwfc = 30  
  ecutrho = 120  
/
&ELECTRONS
/
\end{python}
\begin{itemize}
	\item \verb+ibrav+ selects the type of Bravais unit cell. In the case of silicon, we choose 2, which is the number for face-centered cubic (fcc). In this crystal structure, the lattice vectors are
	\begin{equation}
		{\bf a}_1 = a/2 (-1, 0, 1); \; \; \;
		{\bf a}_2 = a/2 (0, 1, 1); \; \; \;
		{\bf a}_3 = a/2 (-1, 1,0),
	\end{equation}
	where $a$ is the lattice constant given in \verb+celldm(1)+, see also Fig.~\ref{Fig:SiliconBS}, left side. The parameter \verb+celldm(1)+ is given in atomic units, meaning Bohr radii. A full list of possible Bravais lattices can be found in the file \verb+INPUT_PW.txt+.
	\item \verb+nat+ is the total number of atoms per unit cell, \verb+ntyp+ is the total number of {\em different} atoms per unit cell. In fcc silicon, there is only one type (silicon) but two atoms per unit cell.
	\item \verb+occupations+ tells the code how to occupy the computed Kohn-Sham states. \verb+fixed+ means we just occupy all the states below the Fermi level and empty all the states above. Choose this option for insulators; however, for metals (such as graphene in the next section) we need a different option.
	\item Recall that {\sc Quantum ESPRESSO} uses a plane-wave basis to describe the wavefunctions. We should tell the system {\em how many} plane waves should be included in the calculation. This is characterized by a kinetic energy cut-off \verb+ecutwfc+, which implies that we take all the plane waves with momenta such that $\frac{\hbar^2 | {\bf k}|^2}{2m} \leq$ \verb+ecutwfc+. The parameter is given in Rydberg units (1 Ry $= 13.606$ eV). On the Materials Cloud website, see Fig.~\ref{Fig:SSSP}, there is a suggested minimum wavefunction cut-off for every pseudopotential, which for silicon is 30 Ry. Making it larger makes the code slower, but should give a more accurate result.
	\item The code not only stores the plane waves of the electronic states, but also directly the density distribution. Therefore a second cut-off is necessary, \verb+ecutrho+. Because density scales as the square of wavefunctions, and the kinetic energy scales as the square of the momentum, we need to include a density cut-off {\em at least} four times \verb+ecutwfc+. For our case, exactly four times suffices.
	\item The card \verb+&ELECTRONS+ allows us to tell the system how to solve the Kohn-Sham equation. In our case, we just use the default values. However, we need to include the card!
\end{itemize}

The third part of the input contains explicit information about the atoms: their pseudopotentials and positions.
\begin{python}
ATOMIC_SPECIES
  Si    28.086    Si.pbe-n-rrkjus_psl.1.0.0.UPF
ATOMIC_POSITIONS alat
  Si    0.00 0.00 0.00 
  Si    0.25 0.25 0.25 
\end{python}
\begin{itemize}
	\item Below the line \verb+ATOMIC_SPECIES+ you list all the {\em types} of atoms that exist in your unit cell. In our case, it is just silicon. For each type of atom, you give its name (\verb+Si+), its atomic mass in units of u (\verb+28.086+), and the relevant pseudopotential file name.
	\item After \verb+ATOMIC_POSITIONS+ you list the positions of all the atoms. The flag \verb+alat+ means the positions are given in cartesian coordinates in units of the lattice parameter $a$ (\verb+celldm(1)+). Alternatively, one can use \verb+angstrom+ (cartesian coordinates in Angstrom) or \verb+crystal+ (multiples of the primitive lattice vectors). Here we have a silicon atom at the origin, and a second silicon atom at $a (\tfrac{1}{4},\tfrac{1}{4},\tfrac{1}{4} )$.
\end{itemize}

In the final part of the input we tell the code at which momentum points we will do the calculation.
\begin{python}
K_POINTS automatic
  6 6 6 1 1 1
\end{python}
The simplest option is to choose the flag \verb+automatic+, which generates a Monkhorst-Pack grid\cite{Monkhorst:1976cv}. The first three numbers indicate the number of $k$-points in each of the three directions (\verb+ 6 6 6+). The last three provide a possible offset in each direction: \verb+0+ means no offset and thus the inclusion of high-symmetry points like $\Gamma$; \verb+1+ means that you place the momentum points exactly in between the points generated by \verb+0+. A finer momentum mesh is generated if you choose \verb+1+, so that is what we will choose typically.

\subsubsection{Running the code}
Congratulations, you have now written your first DFT input file! To run it, simply use the following command on the command line:
\begin{verbatim}
pw.x -in silicon.scf.in > silicon.scf.out
\end{verbatim} 
The program \verb+pw.x+ is the main component of {\sc Quantum ESPRESSO}. It takes the input file \verb+silicon.scf.in+ and does the self-consistent calculation of the ground state energy and density. On most modern computers, it should not take longer than a few seconds for a system as simple as silicon.

\subsubsection{Reading the output}
The heavy part of the output is sent to the output directory \verb+../out/+. The human-readable part is saved in the file \verb+silicon.scf.out+. Let's read through it together.

The beginning of the output file lists the properties of this calculation, many of them were given by your input file. Some of them were implicit, yet properly picked up. For example, on line 45 you can see that the program will use the PBE exchange-correlation functional. On line 39 we read that we are going to use 4 Kohn-Sham states, with spin degeneracy this corresponds to 8 electrons per unit cell. After that follows properties of the crystal structure, including its symmetries, and the list of momentum points in our grid.

Just before the actual calculation starts, the program estimates the amount of memory needed for this calculation, in the line \verb+Estimated max dynamical RAM per process+. This might be useful to know for more complicated crystal structures, but for silicon this poses no problems.

Between the lines {\em Self-consistent Calculation} and {\em End of self-consistent calculation} the self-consistent DFT loop is iterated until we have reached convergence. After this, the file contains for each momentum point the resulting energies of the Kohn-Sham states. This is directly followed by:
\begin{python}
     highest occupied level (ev):     6.2506

!    total energy              =     -22.83862400 Ry
\end{python}
This is the main output of a DFT calculation: the ground state energy. For clarity, the code also includes the energy of the highest occupied level. This is not exactly the same as the Fermi level, because there might be occupied states with a higher energy at momentum points that were not included in your momentum grid.

\subsection{Homework: Find the lattice constant and bulk modulus of silicon}
{\bf a.} The ground state energy by itself is not a measurable quantity. However, one of the ideas of DFT is that we can find the lattice constant of a crystal by calculating the ground state energy {\em as a function of the input lattice constant}, $E_0(a)$. The predicted actual lattice constant is where this function is minimal. So can you predict with your DFT code the lattice constant of silicon?

{\em Hint: Write a code that automatically generates input files with different values of the lattice constant $a$. Then extract the total energy for every value of $a$.}

{\bf b.} In the previous exercise you calculated the function $E_0(a)$. The change of energy under uniform compression is characterized by the {\em bulk modulus}. What is the value of the bulk modulus you calculated?

\subsection{Bands}
\label{Sec:SiliconBands}

\begin{figure}[t]
\begin{center}
	\includegraphics[width=0.3\textwidth]{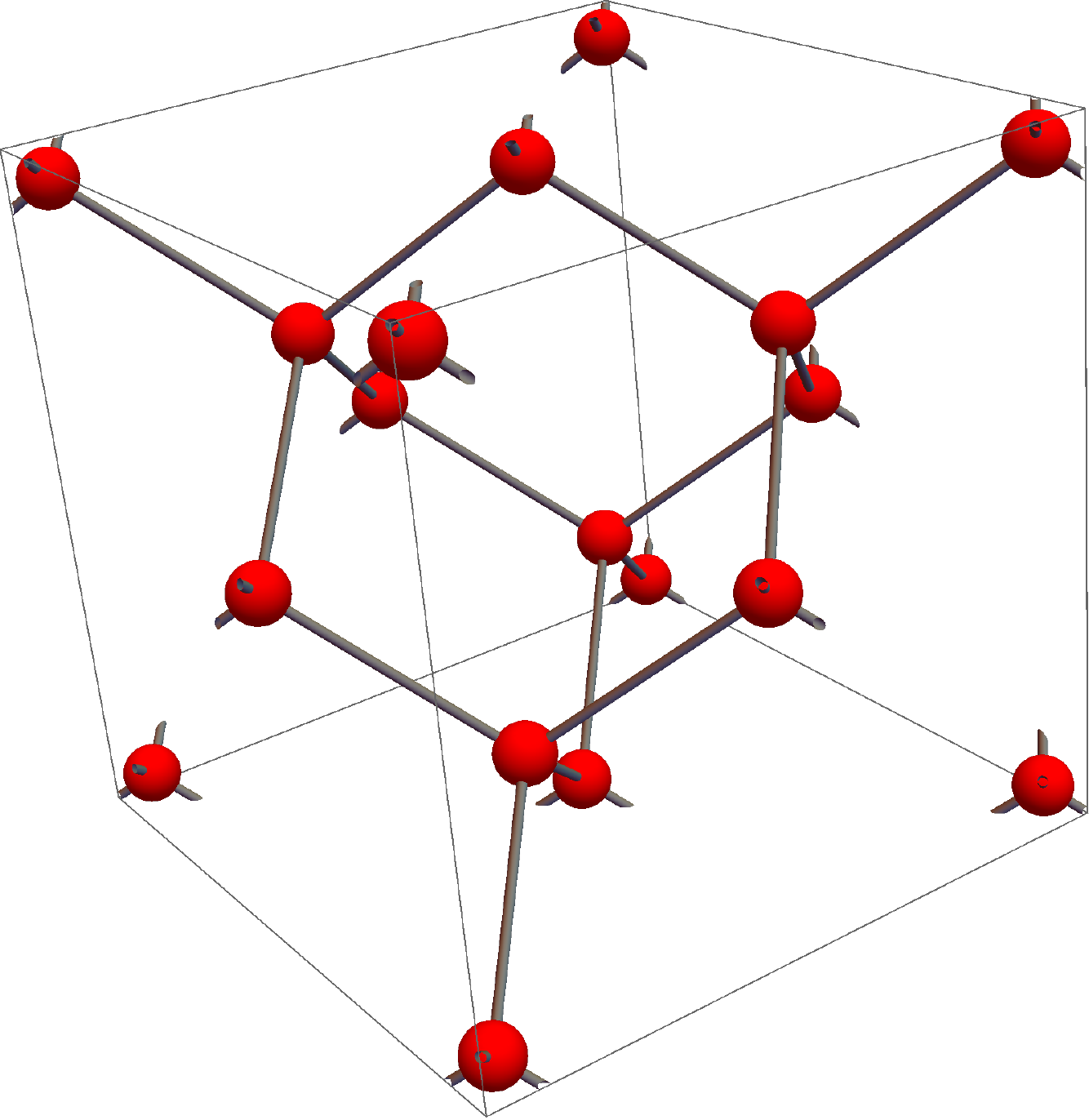}
	\includegraphics[width=0.5\textwidth]{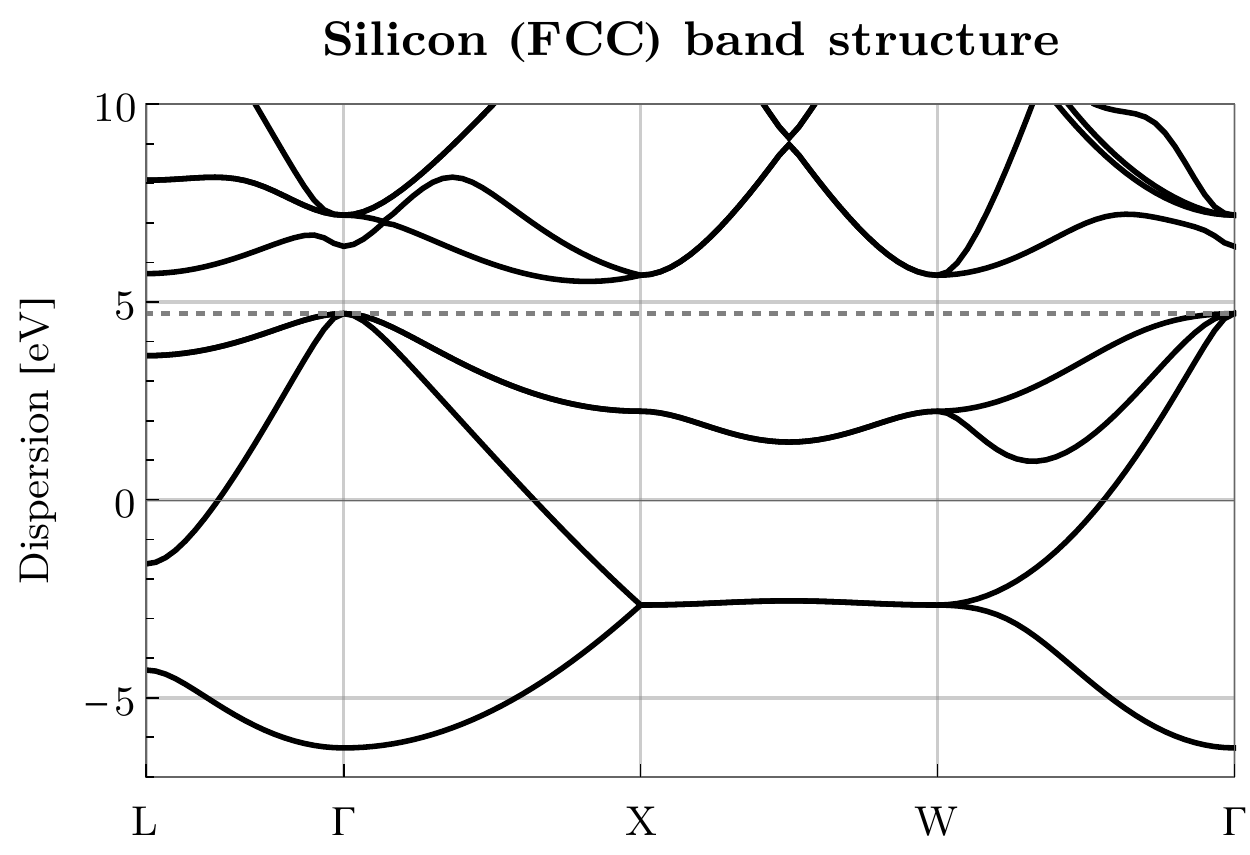}
\end{center}
	\caption{{\bf Left:} Crystal structure of face-centered cubic silicon.
	{\bf Right:} Band structure of silicon as computed using the steps of Sec.~\ref{Sec:SiliconBands}.}
	\label{Fig:SiliconBS}
\end{figure}
Strictly speaking, the Kohn-Sham energies do not correspond to anything physical. However, it turns out that they are a pretty good approximation to the electron band energies of weakly interacting materials. Therefore it is instructive to calculate the Kohn-Sham energies for many points in the Brillouin zone, to get a sense of silicons {\em band structure}. 

A calculation of the bands can only be done after you finished a self-consistent field calculation \makemargintip{with the same parameters}. This is because the bands calculation takes the electronic density (and thus the Kohn-Sham potential) obtained in an \verb+scf+ calculation, and recomputes the Kohn-Sham energies for a new set of chosen momentum points. A good starting point for the bands calculation input file is therefore to copy the \verb+scf+ input file to a new file called \verb+silicon.bands.in+. If you want to write your own bands file from scratch, make sure it has the same \verb+prefix+ as the previous self-consistent field calculation.

At three points, we will change this new input file. First, we need to tell the code that we want to calculate the band structure, so replace the \verb+calculation+ line with this:
\begin{python}
  calculation = 'bands'
\end{python}
For an insulator or semiconductor like silicon, the default setting is that only occupied bands will be computed. If we are interested in the band gap, we also need to calculate the unoccupied bands. To achieve this, add to the \verb+&SYSTEM+ card a line that says we want to calculate 8 bands:
\begin{python}
  nbnd = 8
\end{python}
The final change we need is to the momentum point grid. A customary way to visualize a bandstructure is to choose a {\em path} in the Brillouin zone and compute the bands along this path. This can be done by the command \verb+tpiba_b+, which means that the \verb+K_POINTS+ are in units of $2\pi/ a$. The subscript \verb+_b+ indicates that we can define a path for a bandstructure calculated. For a path L -- $\Gamma$ -- X -- W -- $\Gamma$, we replace the old  \verb+K_POINTS+ card by
\begin{python}
K_POINTS tpiba_b
#  tpiba_b = k-points in units of 2pi/a, in format for band calculation
#  number of k-points (use high-symmetry points only)
 5
#  kx, ky, kz, n. of points between this and next one 
   0.5 0.5 0.5  20
   0.0 0.0 0.0  30
   0.0 0.0 1.0  30
   0.0 1.0 1.0  30
   0.0 0.0 0.0   0
\end{python}
Notice that here we added some comment lines (the one starting with \verb+#+). These will be ignored by the code, and can be useful for our own understanding of the input files. Here, the comments explain us that we have a path with 5 momentum points, and that we have 20 momentum points in between L and $\Gamma$, and so forth.

We can run the bands calculation with this command,
\begin{verbatim}
pw.x -in silicon.bands.in > silicon.bands.out
\end{verbatim} 
It might take a few seconds longer than the \verb+scf+ calculation, because we have more momentum points. 

The output file \verb+silicon.bands.out+ starts out with listing the parameters of the calculation. After the line \verb+Band Structure Calculation+ it will calculate the Kohn-Sham energies of each desired momentum point. At the end of the calculation, you will find lines looking like this:
\begin{python}
     End of band structure calculation

          k = 0.5000 0.5000 0.5000 (   754 PWs)   bands (ev):

    -3.3153  -0.6624   5.1803   5.1803   7.9984   9.7300   9.7300  14.1551

          k = 0.4750 0.4750 0.4750 (   748 PWs)   bands (ev):

    -3.3519  -0.6101   5.1849   5.1849   8.0036   9.7355   9.7355  14.1635
\end{python} 
It signals the end of the calculation, and then it will list for each momentum point all the Kohn-Sham energies in eV. The momentum points themselves are given in units of $\frac{2\pi}{a}$ where $a$ is the lattice constant (\verb+celldm(1)+). 

The {\sc Quantum ESPRESSO} code itself comes with a set of post-processing tools, one of them allows you to plot the band-structure thus calculated. You can also import the output file into your favorite tool (Python, Mathematica, Matlab, GNUplot) and plot it there. The resulting band-structure is shown in Fig.~\ref{Fig:SiliconBS}, right. Notice it is very similar to the actual band-structure!

The band gap calculated using this code is about 0.8 eV, significantly smaller than the actual band gap in silicon of about 1.1 eV. This is common among DFT calculations of semiconductors. On the other hand, many qualitative features -- including the fact that silicon has an indirect band gap -- are reproduced with our simple calculation!

\subsection{Try at home}
Congratulations, you have successfully predicted the properties of a material, purely from first principles! To get comfortable with this technique, try some other three-dimensional semiconductors, and calculate their lattice constant and band-structure.
\begin{enumerate}
	\item First try other zincblende structures like C-diamond, $\beta$-SiC, and GaAs.
	\item Next, calculate a different crystal structure: rock-salt NaCl. Can you find in \verb+PW_input.txt+ how to program its simple cubic structure?
	\item Finally, study hexagonal $\alpha$-SiC. Which one has a lower ground state energy, $\alpha$- or $\beta$-SiC?
\end{enumerate}

\section{Example 2: Graphene}

The study of silicon in the previous section allows you to calculate crystal and electronic band structures of any three-dimensional semiconductor. In the remainder of these notes, we will switch gears and introduce a few new concepts: we discuss how to deal with metals, with two-dimensional materials, and how we can optimize the structure within a unit cell, and how to have more complicated unit cells.

Graphene is a perfect material to do this. It is, as you know, a two-dimensional semi-metal with a hexagonal unit cell.

\subsection{Compute the band-structure}

As with the silicon, we need to write input files for a self-consistent calculation first. Make a new folder \verb+graphene+ and start an input file named \verb+graphene.scf.in+. The first card, \verb+&CONTROL+, is the same as in the silicon case but with only the prefix changed to \verb+graphene+. 

The \verb+&SYSTEM+ card will have some significant changes. Let's write it out in its full totality,
\begin{python}
&SYSTEM
  assume_isolated = '2D'
  ibrav = 4
  celldm(1) = 4.65
  celldm(3) = 6
  nat = 2
  ntyp = 1
  occupations = 'smearing'
  smearing = 'mv'
  degauss =   1.5000000000d-02  
  ecutwfc = 45
  ecutrho = 180
/
&ELECTRONS
/
\end{python}
\begin{itemize}
	\item In the standard DFT implementation, we have periodic boundary conditions in all three direction. The command \verb+assume_isolated = '2D'+ ensures that there is no periodicity (neither in the charge density nor in the Coulomb interactions) in the $z$-direction.
	\item Graphene has an hexagonal lattice, which has \verb+ibrav = 4+. The lattice vectors are given by
	\begin{equation}
		{\bf a}_1 = a (1, 0, 0); \; \; \;
		{\bf a}_2 = a (-\tfrac{1}{2}, \tfrac{\sqrt{3}}{2}, 0); \; \; \;
		{\bf a}_3 = a (0, 0, c/a),
	\end{equation}
	where as before $a=$\verb+celldm(1)+ in Bohr, and $c/a=$\verb+celldm(3)+ is the ratio between the horizontal and vertical lattice size. The value for \verb+celldm(3)+ should be such that, for our 2d set-up in graphene, the vertical unit cell size should be large than the cut-off of the pseudopotentials -- in this case at least 20 Bohr.
	\item In the case of a metal or semimetal, just computing the occupied Kohn-Sham energies is very numerically unstable. Tiny changes can lead to different shapes of the Fermi surface. It is therefore necessary to \makemargintip{smear the occupations} of the Kohn-Sham states, which is ensured by setting \verb+occupations = 'smearing'+. We then also need to set which type of smearing we will use; here we opted for Marzari-Vanderbilt (\verb+mv+) smearing\cite{Marzari:1999gr}, with a width set by \verb+degauss+ in Ry units.
	\item Notice we changed the wavefunction and density cut-offs.
\end{itemize}

\begin{figure}[t]
\begin{center}
	\includegraphics[width=0.35\textwidth]{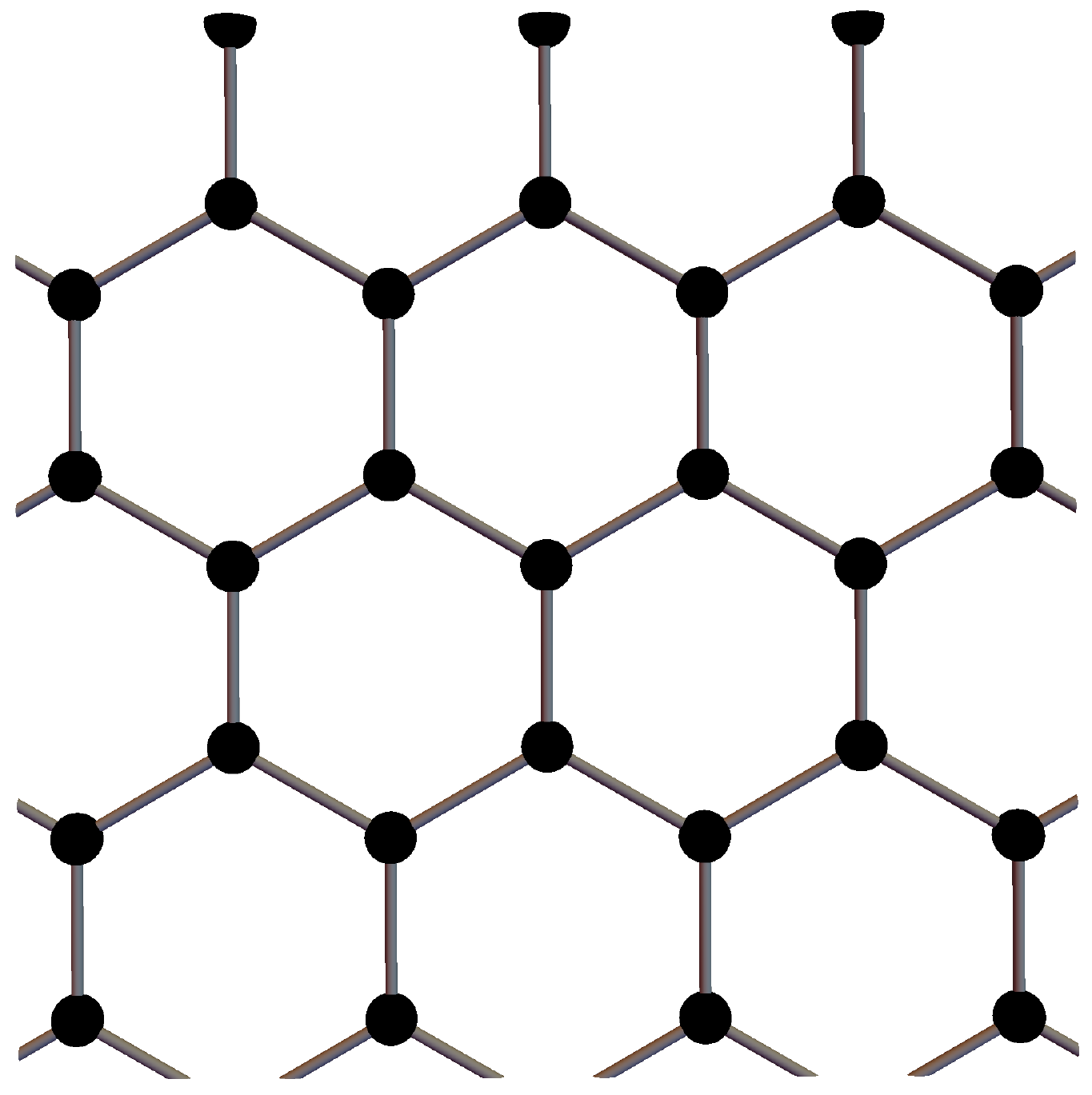}
	\includegraphics[width=0.5\textwidth]{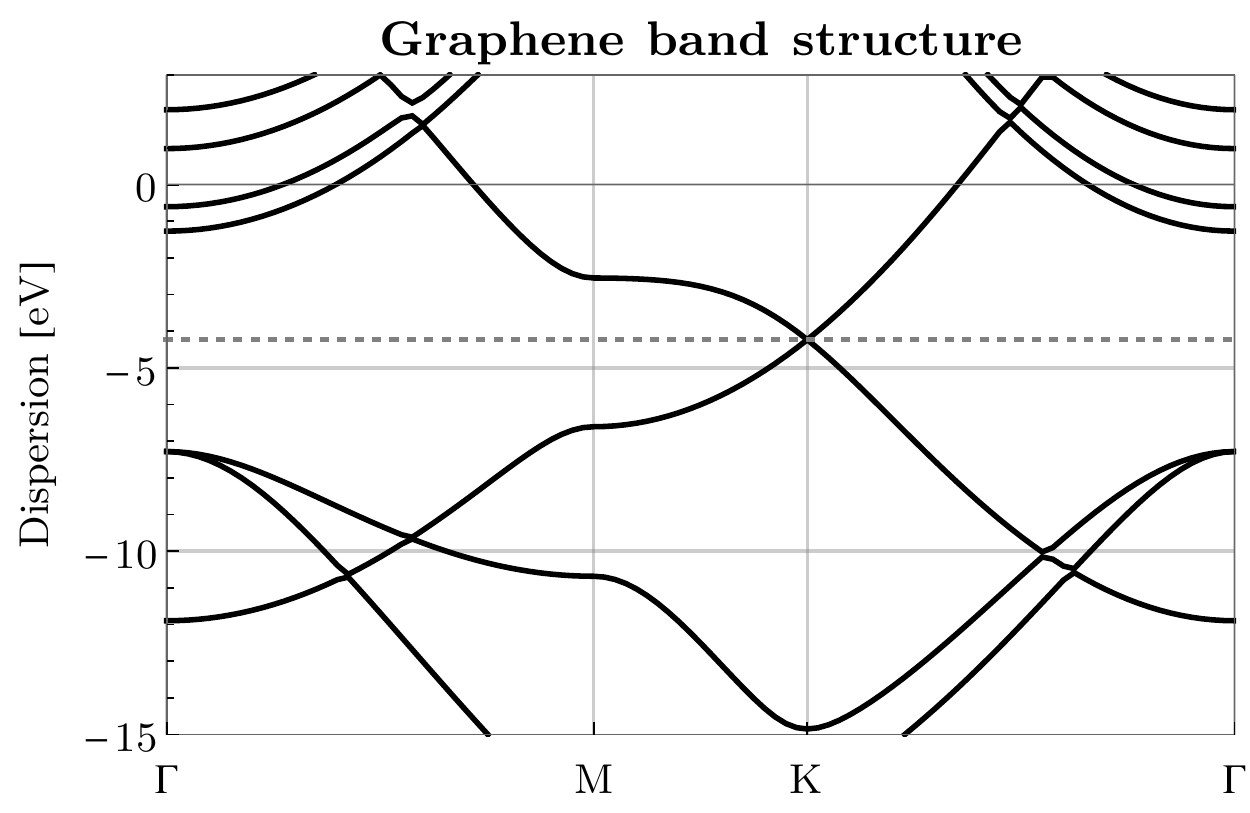}
\end{center}
	\caption{{\bf Left:} The honeycomb crystal structure of graphene.
	{\bf Right:} Band structure of graphene with the Dirac cone clearly visible.}
	\label{Fig:GrapheneBS}
\end{figure}

The last part of the input file tells us where the carbon atoms are going to be, and our choice of momentum points. The only subtleties are in the placement of the carbon atoms, see Fig.~\ref{Fig:GrapheneBS}, left, and in the choice of \verb+K_POINTS+: because we have a two-dimensional system we only need one momentum point in the $z$-direction.
\begin{python}
ATOMIC_SPECIES
  C     12.0107    C.pbe-n-kjpaw_psl.1.0.0.UPF
ATOMIC_POSITIONS alat
  C     0.000000    0.000000    0.000000
  C     0.000000    0.5773503   0.000000
K_POINTS automatic
  9 9 1 1 1 1
\end{python}
Run the self-consistent field calculation by
\begin{verbatim}
pw.x -in graphene.scf.in > graphene.scf.out
\end{verbatim}
The next step is to make our bands calculation input file. Like before, we can just copy the \verb+scf+ input file to a new file \verb+graphene.bands.in+. Make sure you change the type of \verb+calculation+, and the list of \verb+K_POINTS+. For the latter, I suggest a path $\Gamma$ -- M -- K -- $\Gamma$. Because K and M are particularly easily expressed in terms of the reciprocal lattice vectors, we write the \verb+K_POINTS+ in units \verb+crystal_b+:
\begin{python}
K_POINTS crystal_b
 4
   0.000000    0.000000    0.000000     40
   0.500000    0.000000    0.000000     20
   0.333333    0.333333    0.000000     40
   0.000000    0.000000    0.000000     0
\end{python}
In the silicon case we explicitly asked the code to compute more than just the occupied bands, using \verb+nbnd+. Because we are computing a (semi)metal, using the \verb+smearing+ flag, the code automatically calculates some unoccupied bands as well. We do not need to specify the number of bands \verb+nbnd+.

As before, the bands calculation can now be run by the command
\begin{verbatim}
pw.x -in graphene.bands.in > graphene.bands.out
\end{verbatim}
The resulting band-structure, with the characteristic Dirac cone at the Fermi level, can be seen in Fig.~\ref{Fig:GrapheneBS}.

\section{Example 3: WSe$_2$}

\begin{figure}[t]
\begin{center}
	\includegraphics[width=0.48\textwidth]{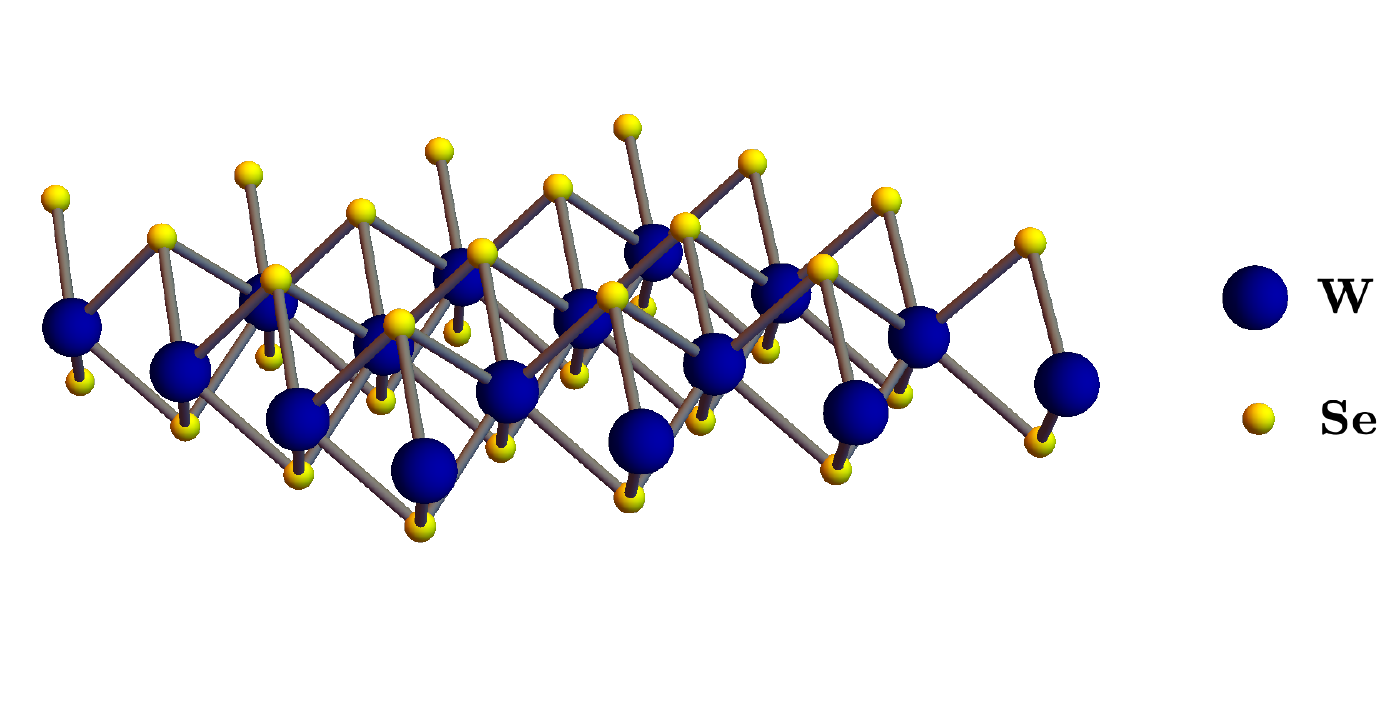}
	\includegraphics[width=0.48\textwidth]{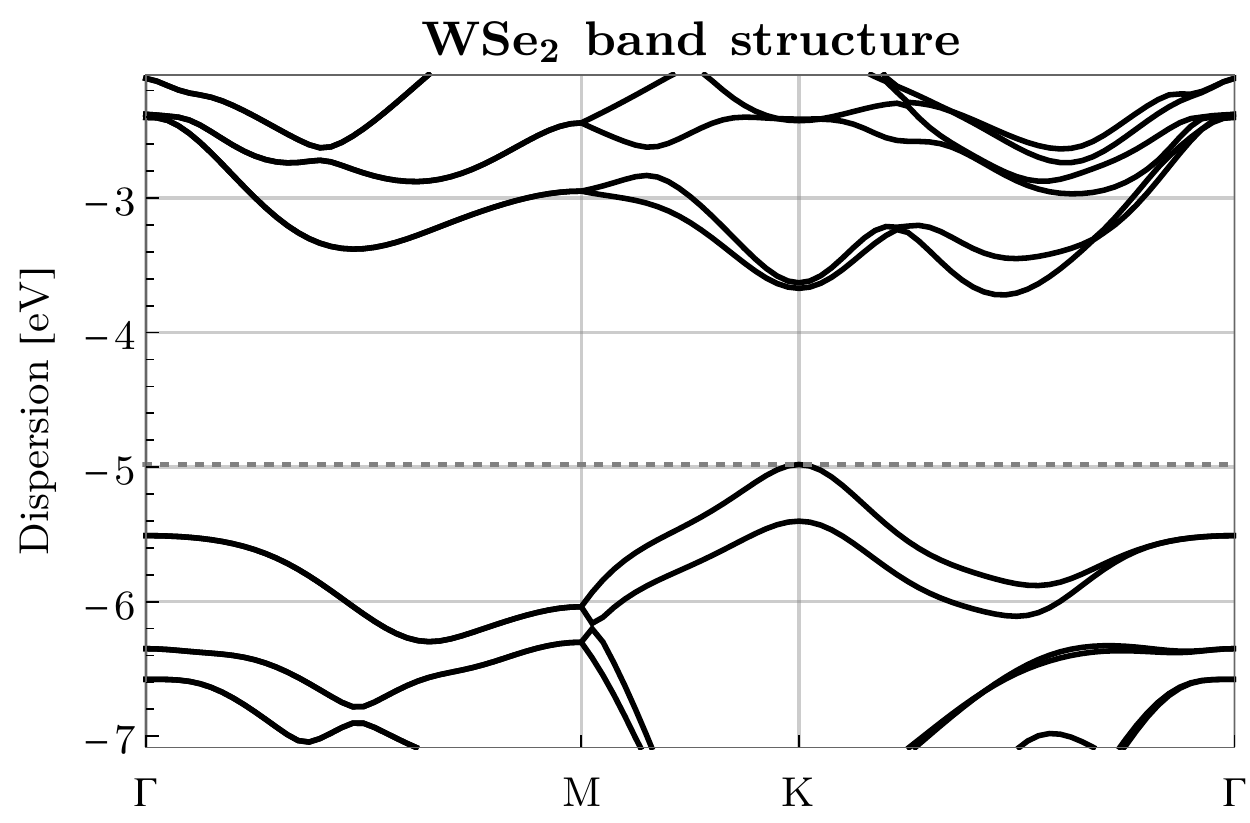}
\end{center}
	\caption{{\bf Left:} Crystal structure of WSe$_2$. 
	{\bf Right:} Final band-structure with spin-orbit coupling, with a sizeable spin-orbit splitting at K of $\Delta^v_{\mathrm{SOC}} = 0.4$ eV.}
	\label{Fig:WSe2}
\end{figure}

The third and final material that allows us to learn some new features of DFT is monolayer WSe$_2$. In-plane it has a honeycomb lattice structure, with on one sublattice the W atoms, and on the other sublattice two Se atoms, displaced in the positive/negative $z$-direction, as shown in Fig.~\ref{Fig:WSe2}, left. Using lattice relaxation calculations, we will be able to find the exact displacement of the Se atoms. Furthermore, WSe$_2$ is a semiconductor with sizeable spin-orbit coupling, and we will show how to include that. 

\subsection{Relax}
In a {\em relaxation} calculation, the DFT code not only computes the ground state energy but also the derivative of the energy with respect to atomic displacements. This corresponds to the forces acting on each atom. If your initial guess of atomic positions is not stable, there will be nonzero forces. The code will suggest a new set of atomic positions based on the direction of those forces. By repeating this until you have no more forces acting on the atoms, you have relaxed the structure and minimized the ground state energy. We will use this feature to calculate the position of the Se atoms in monolayer WSe$_2$.

As before, we start by making a new folder \verb+wse2+ with in there an input file, which we will call \verb+wse2.relax.in+. We will calculate the position of the Se atoms, using a \verb+relax+ calculation. The first card of the input file therefore contains the lines
\begin{python}
  calculation = 'relax'
  prefix = 'wse2'
\end{python}
In the \verb+&CONTROL+ card we can also include a force convergence threshold \verb+forc_conv_thr+, which determines how close to zero we want the final forces to be. In our simple calculation we only need to use the default value, so we do not need include it in our input file.

The remainder of the input file looks like this:
\begin{python}
&SYSTEM
  assume_isolated = '2D'
  ibrav = 0
  nat = 3
  ntyp = 2
  occupations = 'fixed'
  ecutwfc = 30
  ecutrho = 120
/
&ELECTRONS
/
&IONS
/
ATOMIC_SPECIES
  W     183.840    W_pbe_v1.2.uspp.F.UPF
  Se     78.960    Se_pbe_v1.uspp.F.UPF
ATOMIC_POSITIONS angstrom
  W     0.000000    0.000000      0.000000    0   0   0
  Se    0.000000    1.919689645   1.500000    0   0   1
  Se    0.000000    1.919689645  -1.500000    0   0   1
K_POINTS automatic
  8 8 1 1 1 1
CELL_PARAMETERS angstrom
   3.32500000     0.0000000000     0.0000000000
  -1.66250000     2.8795344676     0.0000000000
   0.00000000     0.0000000000    32.0000000000
\end{python}
\begin{itemize}
	\item In the \verb+&SYSTEM+ card, we reverted back to \verb+fixed+ occupations since WSe$_2$ is a semiconductor. Notice how we changed the cut-offs and the number of atoms and types of atoms.
	\item Because we are interested in the atomic positions, it is worthwhile to write out the lattice vectors and the initial atomic positions explicitly in units of Angstrom. We can do so by selecting \verb+ibrav = 0+, meaning we have a free form of the unit cell. We should then explicitly write out the three unit vectors in a new card called \verb+CELL_PARAMETERS+.
	\item We are interested in finding the $z$-position of the Se atoms. We know already that their in-plane coordinates are given by the honeycomb lattice, which are given as the second and third column of the lines after \verb+ATOMIC_POSITIONS angstrom+. In the third column we put the $z$-position. We set W at $z=0$, and we guess an initial distance of the Se atoms at $z = \pm 1.5$ A. The last three numbers indicate \makemargintip{which atomic position coordinates we will relax}. For the Se atoms, \verb+0  0  1+ means we keep the $x,y$ coordinates fixed, and will minimize the ground state energy with respect to the $z$-coordinate. 
	\item We do need to include a new card \verb+&IONS+, where we can specify the properties of the atomic displacements the code will do. Having this card empty just means we choose the default values, but for a \verb+relax+ calculation it has to be there!
\end{itemize}

Run the \verb+pw.x+ code as usual. In the output file \verb+wse2.relax.out+ you can see the two self-consistent loops. Given a set of atomic positions, the ground state energy and forces are calculated. If the forces are larger than the threshold, a new set of atomic positions is proposed after the line saying \verb+ATOMIC_POSITIONS+. After a few iterations, we have reached convergence and we find the following lines containing the {\em final coordinates}:
\begin{python}
Begin final coordinates

ATOMIC_POSITIONS (angstrom)
W        0.000000000   0.000000000   0.000000000    0   0   0
Se       0.000000000   1.919689645   1.678711660    0   0   1
Se       0.000000000   1.919689645  -1.678711660    0   0   1
End final coordinates
\end{python}
We predict the distance between the two Se atoms to be 3.36 $\AA$.

\subsection{Spin-orbit coupling}

In the above lattice relaxation calculation we did not take into account spin-orbit coupling. In general, spin-orbit coupling becomes more important the heavier the element is, which in our case applies to the tungsten (W). Because spin-orbit will likely not influence the position of the Se atoms, we take the atomic positions from the previous calculations and do the standard \verb+scf+ followed by \verb+bands+ to calculate the bandstructure of WSe$_2$. Only this time, we will have spin-orbit coupling.

Make the \verb+wse2.scf.in+ and \verb+wse2.bands.in+ input files. You can combine the elements from the \verb+wse2.relax.in+ and \verb+graphene.bands.in+. Make sure you copy the atomic positions from the previous \verb+relax+ output file into our new input files. To turn on spin-orbit coupling, we need to add the following two lines to the \verb+&SYSTEM+ card:
\begin{python}
  lspinorb = .true.
  noncolin = .true.
\end{python}
The first term turns on spin-orbit coupling, and the second allows for noncollinear spins (so not only up and down but also superpositions). Additionally, we need to have a \makemargintip{fully relativistic pseudopotential to study spin-orbit coupling}. In the W pseudopotential we have used so far, you can find the following line:
\begin{python}
The Pseudo was generated with a Scalar-Relativistic Calculation
\end{python}
Because \verb+Scalar-Relativistic+ implies no spin-orbit coupling, we need to find a new pseudopotential that is fully relativistic! Many different types of pseudopotentials can be downloaded from the {\sc Quantum ESPRESSO} website. Go to \url{https://www.quantum-espresso.org/pseudopotentials/ps-library/}, and download a full relativistic ultra-soft pseudopotential (USPP) for tungsten (W) that works with the PBE functional. After that, update the line in the input files where you give the pseudopotential:
\begin{python}
ATOMIC_SPECIES
  W     183.840    W.rel-pbe-spn-rrkjus_psl.1.0.0.UPF
\end{python}
Finally, the amount of valence electron per W is 14 and per Se is 6, meaning the code computes 26 electrons. If you want also to see the conduction bands, I suggest putting \verb+nbnd = 32+ or higher. 

As before, run the \verb+scf+ first, followed by a \verb+bands+ calculation. You can check in the output files that the number of Kohn-Sham energies is now equal to the number of electrons. Before we included spin-orbit coupling, the spin degeneracy meant we just needed half the amount of Kohn-Sham energies. 

The final band structure is shown in Fig.~\ref{Fig:WSe2}, right. As before, the band gap (here about 1.3 eV) is smaller than experimentally detected (1.7 eV in monolayers). Notably, the valence band at the K point is split due to the spin-orbit coupling, with a splitting of $\Delta^v_{\mathrm{SOC}} = 0.4$ eV, comparable to what is measured in experiments.\cite{Zhang:2016gg}

\section{Further learning}
\label{Sec:Learn}

You now have learned the basics of how to compute crystal structures and electronic bands using density functional theory, implemented in the plane-wave tool {\sc Quantum ESPRESSO}. But there is much more to DFT than just this. There are some additional tools that we haven't discussed, such as the possibility to compute phonon dispersions, Raman or optical spectra, and wannierization. We also haven't looked into modern developments of functionals, such as the inclusion of Van der Waals interactions, hybrid functionals or strong correlations (LDA+U, GW or DFT+DMFT). 

Luckily, there are many online courses available that are more in-depth than this short ToolBox. 
\begin{itemize}
\item The Materials Cloud webpage also contains a set of lectures, including notes, exercises and videos, on how to do DFT with {\sc Quantum ESPRESSO}. You can find them here: \url{https://www.materialscloud.org/learn/}.
\item Many universities have their classes on density functional theory online, for example MIT has \url{https://ocw.mit.edu/courses/materials-science-and-engineering/3-320-atomistic-computer-modeling-of-materials-sma-5107-spring-2005/labs/sections}
\item The source manual of {\sc Quantum ESPRESSO} named \verb+pw_user_guide.pdf+, which comes with downloading the source package, contains a lot of information on what you can do with the code. The source package also contains examples on how to use the code, see the folder \verb+PW/examples+ for input files and ideas for \verb+pw.x+. You can also look at examples of other parts of the code, such as \verb+PHonon/examples+, that show you how to compute phonon dispersions.
\end{itemize}

\section*{Acknowledgements}
These notes would not exist without the enthusiasm of the ToolBoX organizers Jo\~{a}o Ferreira and Michael Sonner. I would also like to thank Marco Gibertini for discussions.

\paragraph{Funding information}
L.R. acknowledges funding from the SNSF in the form of an Ambizione grant.


\nolinenumbers

\end{document}